\documentclass[journal=jctcce,manuscript=article,layout=twocolumn]{achemso}
\setkeys{acs}{articletitle = true}

\usepackage{bm,amsmath}
\usepackage{graphicx}
\usepackage{color}
\usepackage{caption}
\usepackage{subcaption}
\usepackage{booktabs}
\usepackage{footnote}
\usepackage{multirow}
\usepackage[dvipsnames]{xcolor}
\usepackage{algorithmic}
\usepackage[linesnumbered]{algorithm2e}
\usepackage[symbol*]{footmisc}
\usepackage{float}
\usepackage{braket}
\usepackage{cuted}
\usepackage{arydshln}
\usepackage{mathtools}
\usepackage{tabls}
{}
\usepackage[fontsize=10pt]{scrextend}
\usepackage[version=3]{mhchem}
\usepackage{hyperref}
\usepackage{cleveref}
\graphicspath{{./figures/}}

\let\oldnl\nl
\newcommand{\nonl}{\renewcommand{\nl}{\let\nl\oldnl}}

\newtoggle{detailed}
\settoggle{detailed}{false}

\SectionNumbersOn{}

\author{Ryan Stocks}
\affiliation{School of Computing, Australian National University, Canberra, ACT 2601, Australia}
\author{Elise Palethorpe}
\affiliation{School of Computing, Australian National University, Canberra, ACT 2601, Australia}
\author{Giuseppe M. J. Barca}
\affiliation{School of Computing and Information Systems, Melbourne University, Melbourne, VIC 3052, Australia}
\email{giuseppe.barca@unimelb.edu.au}

\title{Multi-GPU RI-HF Energies and Analytic Gradients --- Towards High Throughput \emph{Ab Initio} Molecular Dynamics }

\begin{document}

\begin{abstract}

This article presents an optimized algorithm and implementation for calculating resolution-of-the-identity Hartree-Fock (RI-HF) energies and analytic gradients using multiple Graphics Processing Units (GPUs). The algorithm is especially designed for high throughput \emph{ab initio} molecular dynamics simulations of small and medium size molecules (10-100 atoms). Key innovations of this work include the exploitation of multi-GPU parallelism and a workload balancing scheme that efficiently distributes computational tasks among GPUs. Our implementation also employs techniques for symmetry utilization, integral screening and leveraging sparsity to optimize memory usage and computational efficiency. Computational results show that the implementation achieves significant performance improvements, including over $3\times$ speedups in single GPU AIMD throughput compared to previous GPU-accelerated RI-HF and traditional HF methods. Furthermore, utilizing multiple GPUs can provide super-linear speedup when the additional aggregate GPU memory allows for the storage of decompressed three-center integrals. Additionally, we report strong scaling efficiencies for systems up to 1000 basis functions and demonstrate practical applications through extensive performance benchmarks on up to quadruple-$\zeta$ primary basis sets, achieving floating-point performance of up to 47\% of the theoretical peak on a 4$\times$A100 GPU node.

\end{abstract}

\section{Introduction}

Computing quantum molecular gradients with respect to nuclear movements remains one of the most challenging, important, and computationally demanding applications of quantum chemical theories. These gradients are crucial for identifying equilibrium molecular geometries and transition states.\cite{schaefer_iii_new_1986,schlegel_geometry_2011,pulay_analytical_2014} Additionally, they play a pivotal role in \emph{ab initio} molecular dynamics (AIMD) simulations by directly determining the forces exerted on atoms.\cite{pulay_analytical_2014}

The execution speed of these calculations is a crucial factor in their practical application. It determines the ability to model large molecular systems, to swiftly analyze a vast number of compounds, and to achieve sufficiently long AIMD simulation times.

At the core of electronic structure methods is the iterative Self-Consistent Field (SCF) procedure\cite{szabo_modern_1996}, which is used to solve the Hartree-Fock (HF) and Kohn-Sham (KS) equations\cite{kohn_self-consistent_1965}. The HF method and the SCF form the foundational theoretical and algorithmic infrastructure for KS-Density Functional Theory (KS-DFT) and of more accurate post-HF wave-function methods, such as M\"{o}ller-Plesset perturbation theory\cite{moller_note_1934} and coupled-cluster theories.\cite{szabo_modern_1996}

Due to its strategic importance, tremendous research effort has been dedicated to accelerating HF energy and gradient calculations.

The primary computational bottleneck in the SCF is the construction of the Fock matrix, which necessitates the evaluation of two-electron repulsion integrals (ERIs). The number of ERIs scales as $\mathcal{O}(N^4)$ with system size, posing a significant computational challenge. To mitigate the steep computational cost of the Fock matrix construction, integral screening techniques have been developed to reduce its formal complexity to $\mathcal{O}(N^2)$.\cite{whitten_coulombic_1973,haser_improvements_1989,sabin_molecular_1994,barca_two-electron_2016,barca_three-_2017,thompson_distance-including_2017, black_avoiding_2023}

Further improvements in the Fock matrix assembly time can be achieved through lower scaling algorithms. Specifically, the Coulomb matrix ($\mathbf{J}$) can be obtained in $\mathcal{O}(N\log{N}) $ time using the continuous fast multipole method (CFMM).\cite{white_continuous_1994} The exchange correlation matrix ($\mathbf{K}$) can be obtained in linear time using approaches that integrate optimal screening of its elements, such as LinK,\cite{ochsenfeld_linear_1998} the chain-of-spheres exchange (COSX) method,\cite{neese_efficient_2009} and the seminumerical exchange (sn-K) method.\cite{laqua_highly_2020} Linear scaling in HF calculations is achievable also using molecular fragmentation methods, by neglecting numerically insignificant interactions between distant fragments.\cite{gordon_fragmentation_2012,collins_combined_2014,herbert_fantasy_2019,liu_energy-screened_2019,barca_scaling_2020,barca_scaling_2022}

Despite their lower scaling, which allows access to larger molecular sizes, these approaches have large scaling pre-factors.
This makes them impractical for achieving acceptable time-to-solution in applications requiring low execution latency such as AIMD, where hundreds to millions of these calculations are required for sufficiently long simulation time scales; in interactive AIMD modes, where a human can manipulate molecular geometry on-the-fly while maintaining a smooth and responsive simulation;\cite{luehr_ab_2015,raucci_interactive_2023} or in drug discovery, where conformational searches over libraries containing millions of candidate ligands require each conformer to be quickly geometry-optimized and energetically re-ranked.\cite{habgood_conformational_2020,liu_auto3d_2022,sadybekov_computational_2023}

When targeting large molecular systems, the latency of fragmentation-based calculations can be reduced by using relatively small fragments, while still providing accurate properties,\cite{herbert_fantasy_2019,stocks_high-performance_2024} and by jointly exploiting the large-scale parallelism these methods offer at an algorithmic level.\cite{barca_scaling_2020,pham_development_2020,barca_enabling_2021,barca_scaling_2022,fedorov_multi-level_2023,galvez_vallejo_toward_2023,sattasathuchana_effective_2024}

Thus, while extensive research success has been achieved in enabling HF calculations for larger molecular systems, all aforementioned applications would benefit from a significant reduction in time-to-solution for smaller molecular systems, whether they are fragments in molecular-fragmentation approaches or small and medium sized molecules in non-fragmented approaches.

An effective way to reduce the latency of HF energy and gradient calculations, both for small and large molecules, is to design algorithms that can efficiently leverage the computational power of modern parallel computing architectures such as multi-core CPUs and Graphics Processing Units (GPUs).\cite{ufimtsev_quantum_2008, ufimtsev_quantum_2009, ufimtsev_quantum_2009-1,asadchev_new_2012,liu_new_2014,shan_performance_2014,chow_parallel_2015,chow_scaling_2016,lewis_clustered_2016,kussmann_hybrid_2017,kussmann_employing_2017,huang_accelerating_2018,mironov_efficient_2019,huang_techniques_2020,barca_high-performance_2020,barca_faster_2021,manathunga_harnessing_2021,herault_distributed-memory_2021, seritan_terachem_2021, johnson_multinode_2022,manathunga_quantum_2023,qi_hybrid_2023,williams-young_distributed_2023,asadchev_high-performance_2023,galvez_vallejo_high-performance_2022,hicks_massively_2024,wu_python-based_2024} GPU implementations, in particular, have demonstrated significantly faster SCF calculation times compared to parallel implementations on CPUs.

In this Article, we demonstrate how additional, significant performance benefits can be be obtained for small and medium molecules through a novel approach that efficiently combines the Resolution-of-the-Identity (RI) approximation in Hartree-Fock (RI-HF) with multi-GPU parallelism.

The use of the Coulomb-metric-based RI approximation, also known as density fitting, to accelerate HF calculations was first introduced by Vahtras, Alml{\"o}f, and Feyereisen in 1993.\cite{vahtras_integral_1993} Since then, a substantial body of research has been built upon this method, \cite{fruchtl_implementation_1997,weigend_efficient_2002,neese_improvement_2003,polly_fast_2004,aquilante_low-cost_2007,weigend_hartreefock_2008,sodt_hartree-fock_2008,kossmann_comparison_2009,hohenstein_density_2010,guidon_auxiliary_2010,koppl_parallel_2016,csoka_analytic_2023,bussy_efficient_2024} and it has been clearly established that, when using suitable auxiliary basis sets, the RI-HF absolute energy errors are below 100 $\mu \text{E}_{h}/\text{atom}$, and errors in relative energies and other properties are minimal, typically falling below the basis set incompleteness error.\cite{weigend_fully_2002,weigend_hartreefock_2008}

However, the primary advantage of the RI approach lies in the replacement of the computationally expensive four-electron integrals with simpler two- and three-center integrals. This substitution enables the computational bottlenecks of HF to be reformulated as linear algebra operations that can be executed at near-peak memory or floating-point arithmetic performance. Consequently, the associated computational complexity pre-factors are significantly reduced through more effective hardware utilization. For instance, as detailed in Section \ref{sec:energy_formulation}, the formation of the exchange matrix $\mathbf{K}$ can be implemented using dense matrix-matrix multiplications. This allows the use of vendor-provided GEMM (General Matrix Multiplication) routines, which achieve near-peak double-precision performance.

This advantage is further amplified by the use of GPUs, which have demonstrated superior performance compared to CPUs for linear algebra operations due to their higher memory bandwidth and floating-point throughput. 

Additionally, the replacement of four-center ERIs with three- and two-center analogues is particularly beneficial for GPU execution when high angular momentum basis functions are required. Despite some progress,\cite{asadchev_high-performance_2023,wang_extending_2024} efficient evaluation of four-center ERIs and their gradients involving $f$-type and higher angular momentum atomic orbital functions on GPUs remains an ongoing research challenge, primarily due to the limited number of registers available on the hardware. The RI approximation enables the use of higher angular momentum basis functions while still capitalizing on GPU acceleration, thanks to the simpler, less register-intensive nature of fewer-center ERI calculations. 

To the best of our knowledge, the only GPU-accelerated implementations of RI-HF energy and gradients currently available are in the CP2K and PySCF quantum chemistry software packages.

CP2K provides an RI-HF energy and gradient implementation that utilizes sparsity to make the method cubic rather than quartic scaling~\cite{bussy_sparse_2023}. This implementation is designed to handle large molecular systems with published latencies of the order of minutes to hours. However, we found it significantly underperforms for smaller systems, and therefore it is not included in our comparison.

PySCF~\cite{sun_recent_2020} provides a general and user-friendly interface, but this comes at the cost of computational efficiency, as Python allows for less fine-grained optimization compared to a native C++/CUDA implementation. Recent efforts to accelerate PySCF with the GPU4PySCF package have achieved good performance, particularly for density fitting (RI) calculations.~\cite{wu_python-based_2024} This will serve as our primary performance benchmark, although it currently supports only single-GPU execution. 

The use of multiple GPUs can provide greater efficiency by utilizing the increased aggregate GPU memory, enabling the three center integrals to be distributed between GPUs. This can allow them to be stored decompressed which eliminates the decompression overhead each SCF iteration.

In this work, we provide a detailed formulation of an RI-HF energy and gradient algorithm that efficiently utilizes multi-GPU hardware, while also exploiting available symmetry to enhance performance. The algorithm was implemented in the Extreme-scale Electronic Structure System (EXESS)\cite{galvez_vallejo_toward_2023} and is specifically designed to minimize calculation latency for small and medium sized molecules, enabling high-throughput applications.

The remainder of this Article is structured as follows. Section~\ref{sec:background} details the notation and RI-HF energy and gradient formulations used in the subsequent sections. Section~\ref{sec:algorithms} details the multi-GPU algorithm and implementation details, followed by an analysis of the performance results in Section~\ref{sec:results}.

\section{Theoretical Background}\label{sec:background}

\subsection{Notation}\label{sec:notation}

We use the notation reported in Table~\ref{tab:notation} to represent indices over molecular orbitals and atomic basis functions
\begin{table}[]
\resizebox{\linewidth}{!}{
    \centering
    \begin{tabular}{cl}
    \toprule
        \textbf{Symbol} & \textbf{Representation} \\
    \midrule
        $i, j$ & Doubly occupied molecular orbitals (MOs) \\
        $\mu, \nu, \lambda, \sigma$ & Primary atomic orbital (AO) basis functions $\phi_\mu$ \\
        $P, Q, R, S$ & Auxiliary basis functions $\varphi_P$\\
        $\tilde{x}$ & A batch of index $x$\\
        $N$ & Number of primary basis functions \\
        $N_{occ}$ & Number of occupied orbitals \\
        $N_{aux}$ & Number of auxiliary basis functions \\
    \bottomrule
    \end{tabular}
    \caption{Index symbols}
    \label{tab:notation}
}
\end{table}

All sums are implicitly over the full range of the corresponding index unless otherwise specified. We use chemist's notation for the four center electron repulsion integrals (ERIs)
\begin{equation}
    \begin{aligned}
        (\mu\nu|\lambda\sigma) =& \\
        \int{\phi_\mu}(\boldsymbol{r}_{1})&\phi_\nu(\boldsymbol{r}_1) \frac{1}{|\boldsymbol{r}_1-\boldsymbol{r}_2|}\phi_\lambda(\boldsymbol{r}_2)\phi_\sigma(\boldsymbol{r}_2)\ d\boldsymbol{r}_1d\boldsymbol{r}_2.
    \end{aligned}
\end{equation}

Likewise for the two and three center ERIs
\begin{equation}
        (P|Q) =
        \int{\varphi_P}(\boldsymbol{r}_{1}) \frac{1}{|\boldsymbol{r}_1-\boldsymbol{r}_2|}\varphi_Q(\boldsymbol{r}_2)\ d\boldsymbol{r}_1d\boldsymbol{r}_2,
\end{equation}
and
\begin{equation}
        (\mu\nu|P) =
        \int{\phi_\mu}(\boldsymbol{r}_{1})\phi_\nu(\boldsymbol{r}_1) \frac{1}{|\boldsymbol{r}_1-\boldsymbol{r}_2|}\varphi_P(\boldsymbol{r}_2)\ d\boldsymbol{r}_1d\boldsymbol{r}_2.
\end{equation}

\subsection{RI-HF energy formulation}\label{sec:energy_formulation}

The Hartree-Fock algorithm is an iterative Self Consistent Field (SCF) method in which the minimum energy Slater determinant is found by solving the pseudo-eigenvalue Roothaan equations
\begin{equation}
    \boldsymbol{FC = SC\epsilon},
\end{equation}
where $\boldsymbol{F}$ is the Fock matrix in the AO basis, $\boldsymbol{C}$ are the MO coefficients, $\boldsymbol{S}$ is the AO overlap matrix and $\boldsymbol{\epsilon}$ is a diagonal matrix representing the MO energies. The Fock matrix is itself a function of the MO coefficients so the equations are solved iteratively until convergence. Note that here we will present the formulation using the restricted closed shell approach. Adaptation to unrestricted HF is similar, but requires maintaining two copies of many of the intermediate tensors.

The primary bottleneck of a HF calculation is the Fock build procedure
\begin{equation}\label{fock_build}
    F_{\mu\nu} = h_{\mu\nu} + \sum_{\lambda\sigma}D_{\lambda\sigma}\left((\mu\nu|\lambda\sigma)-\frac{1}{2}(\mu\sigma|\lambda\nu)\right),
\end{equation}
where $h_{\mu\nu}$ is the atomic orbital representation of the core Hamiltonian and $D_{\mu\nu}$ is the doubly occupied density matrix
\begin{equation}
    D_{\mu\nu} = 2\sum_i C_{i\mu}C_{i\nu}.
\end{equation}

The computation of $h_{\mu\nu}$ is relatively cheap and performed only once during the SCF so the calculation of the four-center ERIs, and their ensuing combination with the density matrix to form Fock matrix elements (ERI digestion), constitute the primary bottleneck of the HF calculation. The four center ERIs require too much memory to be stored, and for computational efficiency are re-computed at every iteration of the SCF procedure.

The Resolution of the Identity (RI) approximation~\cite{vahtras_integral_1993} can be used to approximate each of the four-center ERIs in Eq.~\ref{fock_build} as a contraction of two and three-center ERIs
\begin{equation}\label{eq:ri_eri}
    {(\mu\nu|\lambda\sigma)}\approx{(\mu\nu|\lambda\sigma)}_{RI} = \sum_{PQ} (\mu\nu|P) A_{PQ}^{-1} (\lambda\sigma|Q),
\end{equation}
where $A_{PQ} = (P|Q)$ are the auxiliary basis two center ERIs and $A_{PQ}^{-1}$ is its matrix inverse (note that this matrix is often referred to as $\mathbf{J}$ in the literature, however we have renamed it to avoid conflict with the Coulomb operator). This eliminates the four-center ERIs, which are a significant computational burden, and replaces them with combinations of cheaper two and three center ERIs, that can be computed once per SCF and stored in main memory due to their lower space complexity. 


Since $\boldsymbol{A}$ is positive-definite, it is possible to decompose $\boldsymbol{A}^{-1}$ using the Cholesky decomposition
\begin{equation}
    \boldsymbol{A}^L (\boldsymbol{A}^L)^T = \boldsymbol{A}^{-1},
\end{equation}
such that
\begin{equation}\label{eq:symmetric_ri}
    {(\mu\nu|\lambda\sigma)}_{RI} = \sum_{P} B_{\mu\nu}^P B_{\lambda\sigma}^P,
\end{equation}
where
\begin{equation}\label{eq:biap}
    B_{\mu\nu}^P = \sum_{Q} (\mu\nu|Q) A^L_{QP}.
\end{equation}
To maintain numeric stability for larger systems (particularly with the cc-pVTZ-RIFIT auxiliary basis set), we found it crucial to compute the Cholesky decomposition $\boldsymbol{A}^L$ on a single GPU and broadcast it rather than duplicate the computation on each GPU.

The Fock matrix can be conveniently split into three terms
\begin{equation}\label{eq:fock_split}
    F_{\mu\nu} = h_{\mu\nu} + J_{\mu\nu} + K_{\mu\nu},
\end{equation}
where
\begin{equation}\label{eq:j}
    J_{\mu\nu} = \sum_{\lambda\sigma}D_{\lambda\sigma}(\mu\nu|\lambda\sigma),
\end{equation}
is referred to as the Coulomb repulsion term and
\begin{equation}\label{eq:k}
    K_{\mu\nu} = -\frac{1}{2}\sum_{\lambda\sigma}D_{\lambda\sigma}(\mu\sigma|\lambda\nu),
\end{equation}
is the exchange term.

Under the RI approximation, the Coulomb term can be trivially formed using Eq.~\ref{eq:symmetric_ri}
\begin{equation}\label{eq:j_ri}
    J_{\mu\nu} = \sum_{P} B_{\mu\nu}^P \sum_{\lambda\sigma} B_{\lambda\sigma}^P D_{\lambda\sigma},
\end{equation}
which can be computed in $\mathcal{O}(N^2N_{aux})$ operations using two matrix-vector products.

The exchange term is more difficult to compute efficiently. The exchange term can be written as
\begin{align}\label{eq:k_ri_1}
    K_{\mu\nu} &= -\frac{1}{2}\sum_{\lambda\sigma}D_{\lambda\sigma}\sum_{P} B_{\mu\sigma}^P B_{\lambda\nu}^P,
\end{align}
The density matrix can additionally be decomposed into the product of occupied coefficients
\begin{align}\label{eq:k_ri_2}
    K_{\mu\nu} &= -\sum_{i\lambda\sigma}C_{i\lambda}C_{i\sigma}\sum_{P} B_{\mu\sigma}^P B_{\lambda\nu}^P\\
\iftoggle{detailed}{
               &= -\sum_{iP} \left(\sum_{\sigma} B_{\mu\sigma}^P C_{i\sigma} \right) \left( \sum_{\lambda} B_{\lambda\nu}^P C_{i\lambda}\right)\\
               }
               &= -\sum_{iP} B_{i\mu}^{P} B_{i\nu}^{P},
\end{align}
where
\begin{equation}\label{eq:tp}
    B_{i\mu}^{P} = \sum_{\sigma} B_{\mu\sigma}^P C_{i\sigma}.
\end{equation}

This formulation of the exchange term is convenient because it can be computed in $\mathcal{O}(N^2N_{occ}N_{aux})$ operations using two dense matrix-matrix products. Note that this formulation can only be used if the molecular orbital coefficients are known. If only the density is known, then this approach cannot be used. This makes the first SCF iteration significantly slower when using a SAD (Superposition of Atomic Densities) initial guess, though this can be partially overcome by utilizing the sparsity of the SAD density matrix.

\subsection{RI-HF gradient formulation}\label{sec:gradient_formulation}

The RI approximation can also be used to efficiently compute the analytic gradient of the HF energy. The gradient of the HF electronic energy is defined as follows
\begin{align}
\begin{aligned}\label{eq:hf_gradient}
    E_{HF}^\xi = &\sum_{\mu\nu} D_{\mu\nu} h_{\mu\nu}^\xi -  \sum_{\mu\nu}W_{\mu\nu}S_{\mu\nu}^\xi \\&+ \frac{1}{2}\sum_{\mu\nu\lambda\sigma} \left(D_{\mu\nu}D_{\lambda\sigma} - \frac{1}{2}D_{\mu\sigma}D_{\lambda\nu}\right){(\mu\nu|\lambda\sigma)}^\xi,
\end{aligned}
\end{align}
where $\xi$ denotes the derivative of the quantity with respect to a perturbation of the nuclear coordinates, $S_{\mu\nu}^\xi$ is the derivative of the basis function overlap integrals, and
\begin{equation}
    W_{\mu\nu} = 2\sum_{i} C_{i\mu}C_{i\nu}\epsilon_{i},
\end{equation}
is the energy weighted density matrix~\cite{pople_derivative_1979}.

Under the RI approximation,
\begin{align}
\begin{split}
    {(\mu\nu|\lambda\sigma)}^\xi \approx & \frac{d}{d \xi} {(\mu\nu|\lambda\sigma)}_{RI}\\
\end{split}\\
\iftoggle{detailed}{
\begin{split}
    =& \frac{d}{d \xi}\sum_{PQ} (\mu\nu|P) A_{PQ}^{-1} (\lambda\sigma|Q)\\
\end{split}\\
}
\begin{split}
    =\sum_{PQ} &{ (\mu\nu|P)}^\xi A_{PQ}^{-1} (\lambda\sigma|Q) + (\mu\nu|P) A_{PQ}^{-1} {(\lambda\sigma|Q)}^\xi \\
       &- \sum_{PQRS} (\mu\nu|P) A_{PQ}^{-1} A_{QR}^\xi A_{RS}^{-1}(\lambda\sigma|S). \\
\end{split}
\end{align}

Considering just the gradient of the Coulomb term
\begin{align}
    E_C^\xi = &\frac{1}{2}\sum_{\mu\nu\lambda\sigma} D_{\mu\nu}D_{\lambda\sigma} {(\mu\nu|\lambda\sigma)}^\xi\\
\iftoggle{detailed}{
    \begin{split}
    = &\frac{1}{2}\sum_{\mu\nu\lambda\sigma} D_{\mu\nu}D_{\lambda\sigma} \biggl(\sum_{PQ} {(\mu\nu|P)}^\xi A_{PQ}^{-1} (\lambda\sigma|Q) \\
       &- \sum_{PQRS} (\mu\nu|P) A_{PQ}^{-1} A_{QR}^\xi A_{RS}^{-1}(\lambda\sigma|S) \\
       &+ \sum_{PQ} (\mu\nu|P) A_{PQ}^{-1} {(\lambda\sigma|Q)}^\xi\biggr) \\
    \end{split}\\
}{}
    \begin{split}
        = &\sum_{\mu\nu\lambda\sigma} D_{\mu\nu}D_{\lambda\sigma} \biggl(\sum_{PQ} {(\mu\nu|P)}^\xi A_{PQ}^{-1} (\lambda\sigma|Q) \\
      &- \frac{1}{2}\sum_{PQRS} (\mu\nu|P) A_{PQ}^{-1} A_{QR}^\xi A_{RS}^{-1}(\lambda\sigma|S)\biggr) \\
    \end{split}\\
    \begin{split}
    = &\sum_{\mu\nu P} D_{\mu\nu} L^P {(\mu\nu|P)}^\xi
       - \frac{1}{2}\sum_{QR} L^R L^Q A_{QR}^\xi ,
    \end{split}
\end{align}
where
\begin{align}
    L^P &= \sum_{\lambda\sigma Q} D_{\lambda\sigma} A_{PQ}^{-1} (\lambda\sigma|Q)\\
        &=  \sum_Q A^L_{PQ} \sum_{\lambda\sigma} D_{\lambda\sigma}B_{\lambda\sigma}^Q .
\end{align}

Again, the gradient of the exchange term is more expensive
\begin{align}
    E_X^\xi = &-\frac{1}{4}\sum_{\mu\nu\lambda\sigma} D_{\mu\sigma}D_{\lambda\nu} {(\mu\nu|\lambda\sigma)}^\xi\\
\iftoggle{detailed}{
    \begin{split}
    = &-\frac{1}{4}\sum_{\mu\nu\lambda\sigma} D_{\mu\sigma}D_{\lambda\nu} \biggl(\sum_{PQ} {(\mu\nu|P)}^\xi A_{PQ}^{-1} (\lambda\sigma|Q) \\
       &- \sum_{PQRS} (\mu\nu|P) A_{PQ}^{-1} A_{QR}^\xi A_{RS}^{-1}(\lambda\sigma|S) \\
       &+ \sum_{PQ} (\mu\nu|P) A_{PQ}^{-1} {(\lambda\sigma|Q)}^\xi \biggr) \\
    \end{split}\\
}{}
    \begin{split}
    = &-\frac{1}{2}\sum_{\mu\nu\lambda\sigma} D_{\mu\sigma}D_{\lambda\nu} \biggl( \sum_{PQ} {(\mu\nu|P)}^\xi A_{PQ}^{-1} (\lambda\sigma|Q) \\
       &+ \frac{1}{4}\sum_{PQRS} (\mu\nu|P) A_{PQ}^{-1} A_{QR}^\xi A_{RS}^{-1}(\lambda\sigma|S)  \biggr)\\
    \end{split}\\
\iftoggle{detailed}{
    \begin{split}
        =-\frac{1}{2}&\sum_{ij\mu\nu\lambda\sigma} C_{i\mu}C_{i\sigma}C_{j\lambda}C_{j\nu} \biggl( \sum_{PQ} {(\mu\nu|P)}^\xi A_{PQ}^{-1} (\lambda\sigma|Q) \\
       &\frac{1}{4}\sum_{PQRS} (\mu\nu|P) A_{PQ}^{-1} A_{QR}^\xi A_{RS}^{-1}(\lambda\sigma|S)  \biggr)
    \end{split}\\
    \begin{split}
        =&-\frac{1}{2}\sum_{ij} \biggl( \sum_{\mu\nu PQ} {(\mu\nu|P)}^\xi C_{i\mu}C_{j\nu} A_{PQ}^{-1} (ji|Q) \\
       &+ \frac{1}{4}\sum_{PQRS} (ij|P) A_{PQ}^{-1} A_{QR}^\xi A_{RS}^{-1}(ji|S)  \biggr)
    \end{split}\\
}{}
    \begin{split}
        = & - \frac{1}{2} \sum_{ij\mu\nu P} {(\mu\nu|P)}^\xi C_{i\mu}C_{j\nu} X_{ij}^P
          + \frac{1}{4}\sum_{ijQR} X_{ij}^Q A_{QR}^\xi X_{ij}^R,
    \end{split}
\end{align}
where
\begin{align}
    X_{ij}^P &= \sum_{Q} A_{PQ}^{-1} (ij|Q)\\
             &= \sum_{Q} A_{PQ}^L B_{ij}^Q.
\end{align}

The transformation to the occupied orbital basis reduces the cost of communicating the $X_{ij}^P$ tensor between GPUs which will be explained further in Section~\ref{sec:alg_gradient}.

The full RI-HF gradient becomes
\begin{align}
\begin{aligned}\label{eq:rihf_gradient}
    E_{HF}^\xi = &\sum_{\mu\nu} D_{\mu\nu} h_{\mu\nu}^\xi -  \sum_{\mu\nu}W_{\mu\nu}S_{\mu\nu}^\xi \\&+ \sum_{\mu\nu P} Y_{\mu\nu}^P {(\mu\nu|P)}^\xi - \frac{1}{2}\sum_{PQ} Z_{PQ} A_{PQ}^\xi,
\end{aligned}
\end{align}
where
\begin{equation}
    Y_{\mu\nu}^P = D_{\mu\nu} L^P - \frac{1}{2} \sum_{ij} C_{i\mu}C_{j\nu} X_{ij}^P,
\end{equation}
and
\begin{equation}
    Z_{PQ} = L^P L^Q - \frac{1}{2}\sum_{ij} X_{ij}^P X_{ij}^Q.
\end{equation}

This formulation enables the coefficients for the two and three center ERI derivatives ($Z_{PQ}$ and $Y_{\mu\nu}^P$, respectively) to be computed first, and then accumulated directly into the final gradient while computing $(\mu\nu|P)^\xi$ on the fly.

\subsection{Algorithmic challenges}

Here we will briefly summarise the challenges associated with an efficient RI-HF energy and gradient implementation on multiple GPUs.
\begin{itemize}
    \item {\textbf{Large intermediate tensors}. The $B_{\mu\nu}^P$ matrix scales cubicly with the system size which can quickly saturate the small GPU memory. Ideally this matrix must be distributed between multiple GPUs rather than duplicated or stored on the host. There is significant predictable sparsity that can also be exploited as the three center ERIs $(\mu\nu|P)$ are numerically significant only if the basis functions $\phi_\mu$ and $\phi_\nu$ sufficiently overlap.}
    \item{\textbf{Communication overheads}. Transfer of data to the host CPU and between GPUs is required for control flow and to minimize work duplication, but is significantly slower than the memory bandwidth and compute throughput within a GPU. For example an A100 GPU can perform 19.5 TFLOP/s whilst the peak NVLINK inter-GPU bandwidth is 600 GB/s so transferring 1 double precision value is equivalent to 260 floating point operations. Thus data communication must be minimized. }
    \item{\textbf{Parallel efficiency}. To minimize parallel time-to-solution, it is important that work is efficiently distributed between the GPUs and that non-parallelizable overhead (such as control flow or single GPU operations) is minimized or overlapped with parallelizable work. }
    \item{\textbf{Symmetry utilization}. Many intermediates are symmetric. For example, $B_{\mu\nu}^P$ is symmetric with respect to exchange of $\mu$ and $\nu$. This should be exploited to minimize the computational work and memory footprint, although there are limited options in current BLAS libraries for utilizing these properties.}
\end{itemize}

\section{\label{sec:algorithms}Algorithm and implementation}

In this section, we present an efficient multi-GPU algorithm for the RI-HF energy and gradients, which addresses the aforementioned algorithmic challenges.  

\subsection{Energy}

The overarching control flow of the SCF algorithm is the same as a standard GPU accelerated Hartree-Fock, however the bottleneck Fock build procedure at each iteration is entirely replaced with an initial computation of the two and three center ERIs followed by an RI Fock build at each iteration consisting of dense linear algebra. DIIS convergence acceleration is utilized to minimize SCF iterations. The Fock matrix extrapolation and diagonalization is performed on a single GPU between each iteration, and the resulting coefficients are broadcast to the other GPUs. The RI specific components of the Hartree-Fock procedure are shown in Algorithm~\ref{alg:rihf_energy}.

The $B_{\mu\nu}^P$ tensor is computed once at the beginning of the calculation (lines \ref{alg:ln:begin_B}-\ref{alg:ln:end_B} of Algorithm~\ref{alg:rihf_energy}) and reused for each iteration of the SCF procedure (lines \ref{alg:ln:begin_scf}-\ref{alg:ln:end_scf}. This tensor is statically distributed between GPUs with each GPU storing a subset (batch) of the auxiliary index $P$. This distribution reduces the memory footprint on each GPU to $\mathcal{O}(\frac{N^2N_{aux}}{\#\ GPUs})$ which enables larger systems to be considered on multiple GPUs by utilizing the aggregate GPU memory across the node. Host-device transfer of this matrix would be a significant bottleneck for all system sizes reachable on current hardware so the three center ERI calculations are duplicated on each GPU to enable immediate contraction with the corresponding subset of the $A_{PQ}^L$ matrix.
These three center ERIs are computed and contracted into the GPU local $B_{(\mu\nu)}^P$ matrix in batches to minimize memory consumption. Only significant shell pairs $(\mu\nu)$ are considered at this stage as described further in Section~\ref{sec:sparsity}

\begin{algorithm}[!htb]
    \caption{Multi-GPU calculation of the RI-HF energy.}\label{alg:rihf_energy}
    Calculate $h_{\mu\nu}$ and $A_{PQ}^{L}$
    
    Statically assign subset $\tilde{P}$ to each GPU

    \ForAll{batch $\tilde{Q}$}{\label{alg:ln:begin_B}
        Calculate $(\mu\nu|\tilde{Q})$

        $B_{(\mu\nu)}^{\tilde{P}} = \sum_{\tilde{Q}} (\mu\nu|\tilde{Q}) A_{\tilde{P}\tilde{Q}}^L$
    \label{alg:ln:end_B}

    }
    \If{sufficient GPU memory for $B_{\mu\nu}^P$}{ \label{alg:ln:begin_decompress}
            Decompress $B_{\mu\nu}^{\tilde{P}} = B_{(\mu\nu)}^{\tilde{P}}$
    }\label{alg:ln:end_decompress}
    \While{SCF not converged} {\label{alg:ln:begin_scf}
        $F_{\mu\nu} = h_{\mu\nu}$
        
        \ForAll{sub-batches $\tilde{P}$}{
        
            Decompress $B_{\mu\nu}^{\tilde{P}} = B_{(\mu\nu)}^{\tilde{P}}$ if required

            $D_{\mu\nu} = 2 \sum_i C_{i\mu} C_{i\nu}$

            $M^{\tilde{P}} = \sum_{\mu\nu} B_{\mu\nu}^{\tilde{P}} D_{\mu\nu}$\label{alg:ln:begin_f}
            
            $B_{i\mu}^{\tilde{P}} = \sum_{\nu} B_{\mu\nu}^{\tilde{P}} C_{i\nu} $

            $F_{\mu\nu} \mathrel{+}= \sum_{\tilde{P}} M^{\tilde{P}}B_{\mu\nu}^{\tilde{P}} - \sum_{i\tilde{P}} B_{i\mu} ^{\tilde{P}} B_{i\nu}^{\tilde{P}}$\label{alg:ln:end_f}
        }
        Extrapolate and diagonalize $F_{\mu\nu}$ to obtain new trial $C_{i\mu}$\label{alg:ln:diag}
    }\label{alg:ln:end_scf}
\end{algorithm}

Following the computation of $B_{(\mu\nu)}^P$, if there is sufficient GPU memory available, it is best to decompress the matrix to no longer utilize the shell pair sparsity to avoid decompression at every stage of the SCF procedure (lines \ref{alg:ln:begin_decompress}-\ref{alg:ln:end_decompress}). Then at each iteration of the SCF, the Fock matrix $F_{\mu\nu}$ is built using a sequence of dense matrix-matrix and matrix-vector multiplications (lines \ref{alg:ln:begin_f}-\ref{alg:ln:end_f}) and used to generate a new trial set of MO coefficients $C_{i\mu}$ (line \ref{alg:ln:diag}).

\subsubsection{Basis function pair sparsity}\label{sec:sparsity}

\begin{figure}
    \centering
    \includegraphics[width=1\linewidth]{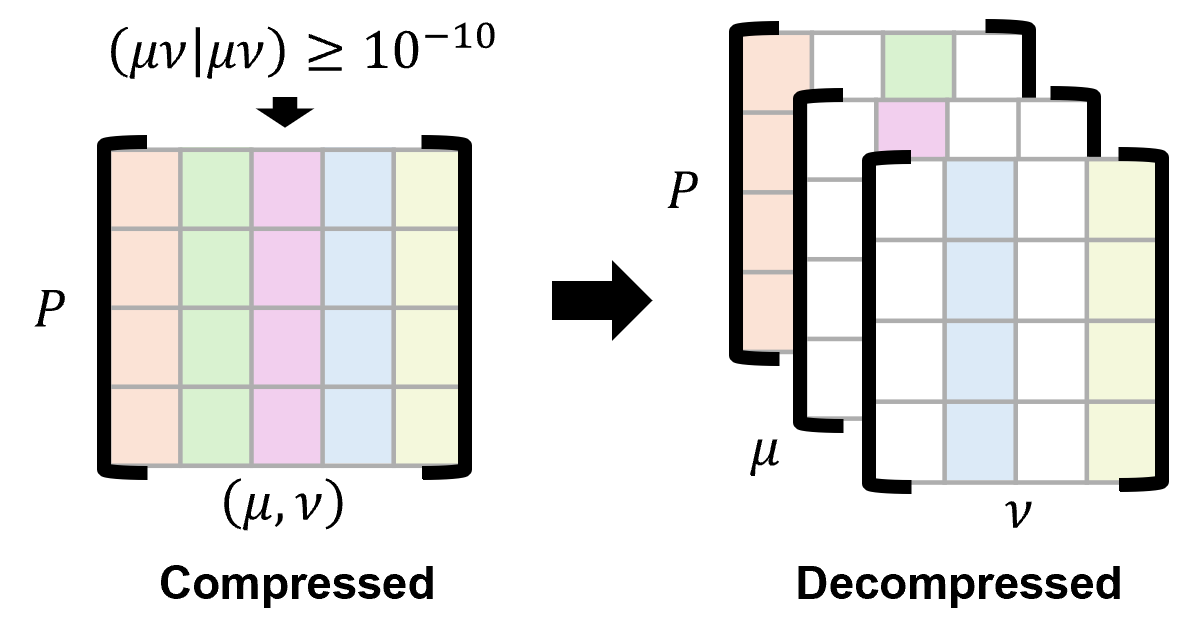}
    \caption{Illustration of basis function pair screening and decompression.}
    \label{fig:compression}
\end{figure}

The primary bottleneck to the size of system that can be considered with a RI-HF calculation is the storage of the $B_{\mu\nu}^P$ tensor. However this tensor has significant sparsity that can be exploited to expand the range of systems that can be considered. The Schwarz inequality~\cite{whitten_coulombic_1973} bounds the magnitude of a four center repulsion integral 
\begin{equation}
     (\mu\nu|\lambda\sigma) \leq \sqrt{(\mu\nu|\mu\nu)}\sqrt{(\lambda\sigma|\lambda\sigma)}.
\end{equation}
Hence shell pairs $(\mu\nu)$ can be neglected entirely if $(\mu\nu|\mu\nu)$ is sufficiently small ($\leq 10^{-10}$). This reduces the cost of computing the three center ERIs and formation of $B_{\mu\nu}^P$ by the fraction of screened out shell pairs. This also allows $B_{\mu\nu}^P$ to be stored in a compressed format with only significant shell pairs and can be decompressed on the fly when required as illustrated in Figure~\ref{fig:compression}.

This decompression is inefficient on GPUs due to the large number of uncoalesced memory operations and does not utilise all sparsity of the $\boldsymbol{B}$ matrix. The GPU architecture is extremely sensitive to uncoalesced memory operations (consecutive GPU threads accessing non-contiguous data in memory) which will occur in any decompression strategy with granularity finer than the number of threads in a warp. If there is sufficient GPU memory, it is most efficient to decompress $\boldsymbol{B}$ immediately following its computation rather than every SCF iteration. It would be interesting to compare additional compression methods in future work that may enable larger compression ratios or more efficient decompression.

\subsubsection{Symmetry utilization}

Many of the intermediate tensors ($\boldsymbol{A}, \boldsymbol{B}, \boldsymbol{D}, \boldsymbol{F}, \boldsymbol{J}, \boldsymbol{K}$) are symmetric with respect to the exchange of atomic orbital indices. Hence, forming them using naive dense matrix multiplications performs twice as many floating point operations as is strictly necessary. Unfortunately, there is no vendor provided BLAS operation for a matrix multiplication where the result is symmetric, so a custom symmetric GEMM routine was implemented.

\begin{figure}
    \centering
    \includegraphics[width=1\linewidth]{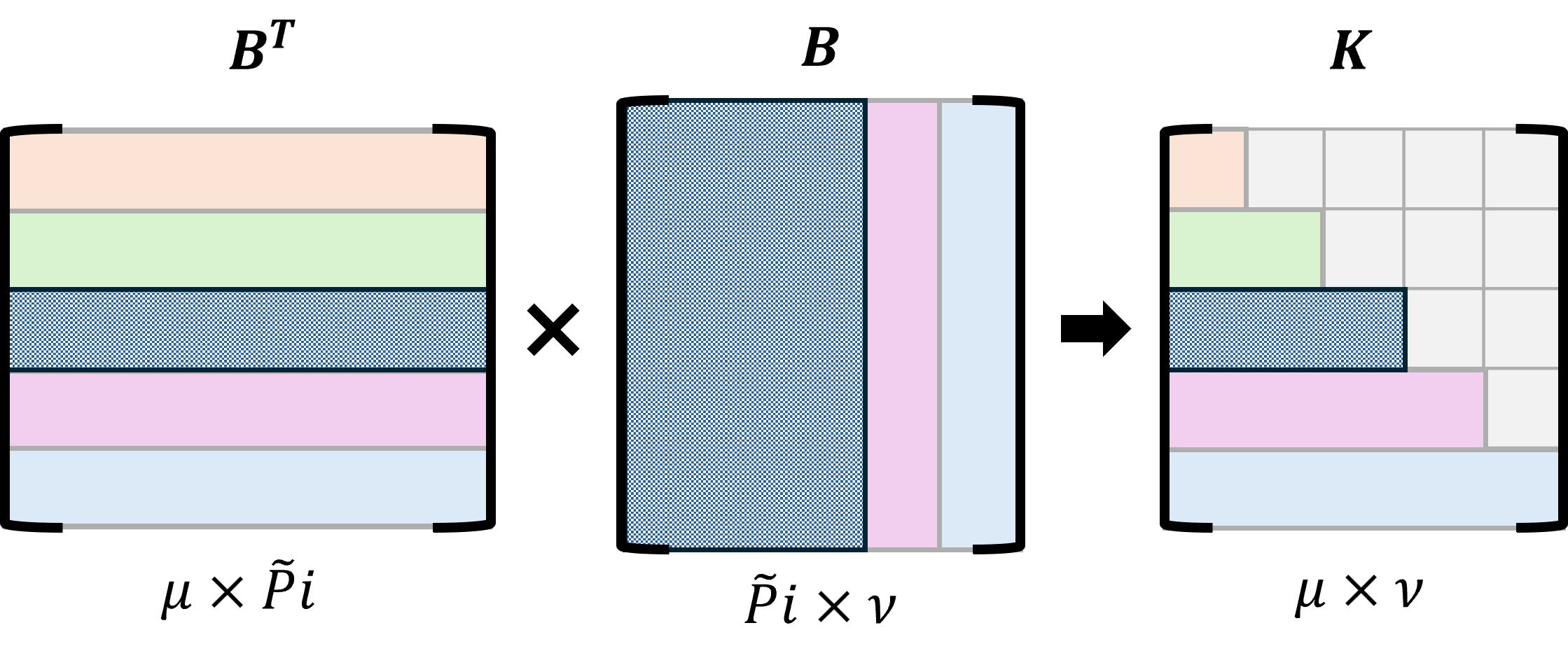}
    \caption{Example of symmetry utilization to reduce number of floating point operations for the formation of $K_{\mu\nu}=\sum_{i\tilde{P}} B_{i\mu}^{\tilde{P}} B_{i\nu}^{\tilde{P}}$ by forming only the blocked lower triangle.}
    \label{fig:symmetric_gemm}
\end{figure}

An example for the formation of the lower triangle of $\boldsymbol{K}$ is shown in Figure~\ref{fig:symmetric_gemm}. The matrix multiplication is split into batches by row of the resultant matrix. Each of these batches can be performed in a separate GPU stream to maximise resource utilization in case a single batch does not fully utilize the hardware. In future, it could be beneficial to implement a single matrix multiplication kernel that utilizes the symmetry to reduce kernel call overhead, however this will require significant tuning and will be less portable than the batched GEMM approach. For small matrices it is more efficient to use a single matrix multiplication than batching multiple smaller multiplications.

\subsection{Gradient}\label{sec:alg_gradient}

Following the evaluation of the RI-HF energy, the gradient calculation can reuse the $B_{\mu\nu}^P$ and $A_{PQ}^L$ intermediates to avoid recomputation of the two and three center ERIs. The gradients benefit significantly from the use of the RI approximation as the bottleneck becomes the computation of the three center integral gradients $\frac{d}{dx} (\mu\nu|P)$ rather than the four center integral gradients $\frac{d}{dx} (\mu\nu|\lambda\sigma)$. Algorithm~\ref{alg:rihf_gradient} shows the full sequence of communication and linear algebra routines for the calculation of the RI-HF gradients.

\begin{algorithm}[!htb]
    \caption{Multi-GPU calculation of the RI-HF energy gradient.}\label{alg:rihf_gradient}
    Use same subset $\tilde{P}$ used for energy calc

    $X_{ij}^R = 0$ \label{alg:ln:begin_bij}
    
    \ForAll{sub-batches of $\tilde{P}$}{
    
        Decompress $B_{\mu\nu}^{\tilde{P}} = B_{(\mu\nu)}^{\tilde{P}}$ if required

        $B_{ij}^{\tilde{P}} = \sum_{\mu\nu} C_{i\mu}C_{j\nu} B_{\mu\nu}^{\tilde{P}}$
    }\label{alg:ln:end_bij}
    
    Gather $B_{ij}^R$ on host and broadcast to GPUs \label{alg:ln:bij_broadcast}
    
    $E^\xi_{HF} = 0$
    
    \If{GPU 0} {\label{alg:ln:begin_onee}
        $E^\xi_{HF} \mathrel{=} \sum_{\mu\nu} D_{\mu\nu}h^\xi_{\mu\nu} + W_{\mu\nu}S^\xi_{\mu\nu}$
    
        $X_{ij}^{R} \mathrel{+}= \sum_{P} B_{ij}^{P} A_{PR}^{L}$
        
        $L^R = \sum_i X_{ii}^R$
    
        $Z_{PQ} = L^P L^Q - \frac{1}{2} \sum_{ij} X_{ij}^P X_{ij}^Q$
        
        $E^\xi_{HF} \mathrel{+=} -\frac{1}{2}\sum_{PQ} Z_{PQ} (P|Q)^\xi$
    }\label{alg:ln:end_onee}
    \ForAll{batch $\tilde{Q}$ dynamically distributed to GPUs}{\label{alg:ln:begin_3c}
    
        $X_{ij}^{\tilde{Q}} \mathrel{+}= \sum_{P} B_{ij}^{P} A_{P\tilde{Q}}^{L}$
        
        $L^{\tilde{Q}} = \sum_i X_{ii}^{\tilde{Q}}$

        $Y_{\mu\nu}^{\tilde{Q}} = D_{uv} L^{\tilde{Q}} - \frac{1}{2} \sum_{ij} C_{i\mu}C_{j\nu} X_{ij}^{\tilde{Q}}$
    
        $E^\xi_{HF} \mathrel{+=} \sum_{\mu\nu\tilde{Q}} Y_{\mu\nu}^{\tilde{Q}} (\mu\nu|\tilde{Q})^\xi$
    }\label{alg:ln:end_3c}
    
    Reduce $E^\xi_{HF}$ on host\label{alg:ln:final_grad_reduction}
    
\end{algorithm}

The primary challenge of the RI-HF gradients is that all elements of the $B_{\mu\nu}^P$ matrix are required on each GPU. These communication costs are minimized by transforming $B_{\mu\nu}^P$ to the molecular orbital basis $B_{ij}^R$ (lines \ref{alg:ln:begin_bij}-\ref{alg:ln:end_bij} of Algorithm \ref{alg:rihf_gradient}). Since only the occupied orbitals are required, this has a significantly lower memory footprint than the full atomic orbital representation ($\approx 5\%$ with double zeta basis sets and $<1\%$ with triple zeta basis sets for the glycine systems considered).

Following the broadcast of $B_{ij}^R$ (line \ref{alg:ln:bij_broadcast}), a single GPU computes the one electron and two center ERI contributions to the gradient (lines \ref{alg:ln:begin_onee}-\ref{alg:ln:end_onee}). The one electron component can be overlapped with the broadcast of $B_{ij}^R$ as it relies only on the MO coefficients. Following the broadcast, $Z_{PQ}$ is calculated and contracted with the two center ERI gradients into the final gradient.

Each worker GPU is then dynamically distributed a batch of the auxiliary basis functions $\tilde{Q}$ for which to calculate the three center ERI contribution to the gradient (lines \ref{alg:ln:begin_3c}-\ref{alg:ln:end_3c}. The first GPU joins the worker pool for the three center component following the computation of the one electron and two center ERI contributions to the gradient. Finally the GPU-local gradients are reduced on the host CPU to form the final gradient (line~\ref{alg:ln:final_grad_reduction}).

\subsection{\textit{Ab Initio} Molecular Dynamics}

In addition to the RI-HF energy and gradients, we have implemented a lightweight wrapper to demonstrate the utility of the implementation for ~\textit{ab initio} molecular dynamics (AIMD). The wrapper performs Born-Oppenheimer molecular dynamics using a Verlet integration scheme to update the positions and velocities of the nuclei each time step. In this work we simulate in the NVE micro-canonical ensemble to demonstrate energy conservation. We additionally re-use the converged orbitals from the previous time step as the initial guess for the next time step to reduce number of SCF iterations. The orbital initial guess also avoids the additional cost of the RI Fock build from a guess density at the first SCF iteration each time step.

\section{\label{sec:results} 
Results}

All calculations were performed on a single node of the NERSC Perlmutter supercomputer. Each node has a single AMD EPYC 7763 CPU and 4 NVIDIA A100 40GB GPUs. All inputs and outputs as well as the scripts for data analysis and plot creation are provided as Supporting Information.

We have elected to use a variety of common basis sets, in particular cc-pVDZ~\cite{dunning_gaussian_1989}, def2-SVP~\cite{weigend_balanced_2005} and 6-31G**~\cite{hariharan_influence_1973}, along with the triple zeta basis sets cc-pVTZ~\cite{dunning_gaussian_1989} and def2-TZVP~\cite{weigend_balanced_2005}. We additionally perform a calculations on a 10 water cluster with def2-QZVP~\cite{weigend_gaussian_2003} to demonstrate the \textit{g} function capability of the RI-HF implementation. Each of the primary basis sets is paired with the corresponding RI auxiliary basis set, namely cc-pVDZ-RIFIT~\cite{weigend_efficient_2002}, def2-SVP-RIFIT~\cite{weigend_ri-mp2_1998}, 6-31G**-RIFIT~\cite{tanaka_optimization_2013}, cc-pVTZ-RIFIT~\cite{weigend_efficient_2002}, def2-TZVP-RIFIT~\cite{weigend_ri-mp2_1998}, and def2-QZVP-RIFIT~\cite{hattig_optimization_2005} respectively. All basis sets were obtained from Basis Set Exchange~\cite{pritchard_new_2019} and used in Cartesian form.

The majority of the performance tests are run on poly-glycine chains of varying lengths for consistency with our RI-MP2 gradients work~\cite{stocks_high-performance_2024}. Each glycine (\ce{C2 H3 N O}) adds 30 electrons across 7 atoms. We denote a poly-glycine chain of length $n$ as gly\textsubscript{n} \ce{=H(C2 H3 N O)_n OH}. With double zeta basis sets this has a maximum of \emph{d} orbital angular momentum in the primary basis and \emph{f} angular momentum in the auxiliary basis, whilst with triple zeta basis sets it has a maximum of \emph{f} and \emph{g} angular momentum for the primary and auxiliary basis sets respectively. The floating point throughput where reported is calculated by counting floating point operations performed in DGEMM routines divided by the total execution time. Floating point operations performed outside DGEMMs (in the ERIs, eigensolver, etc.) are not counted as they are difficult to estimate at runtime and constitute a small portion of the overall FLOP count.

The triple and quadruple zeta tests were conducted on Perlmutter nodes with 80 GB GPUs due to the higher memory footprint.

\subsection{Energy Conservation}

\begin{figure}[!h]
    \centering
    \includegraphics[width=1\linewidth]{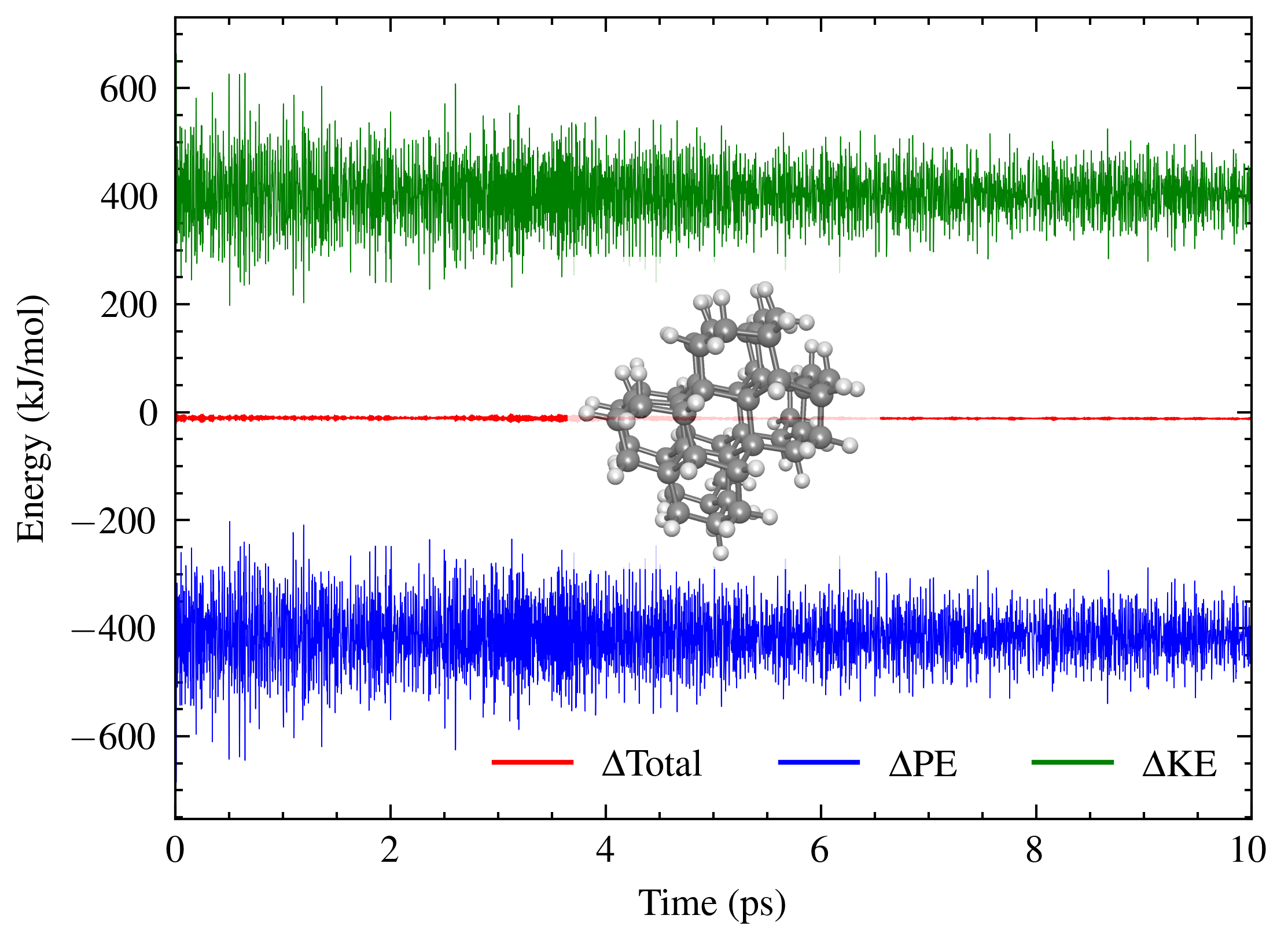}
    \caption{Energy conservation over a 10 ps RI-HF simulation of a \ce{C54H56} fully hydrogenated diamond nanoparticle with the def2-SVP basis set on 4 A100 GPUs}
    \label{fig:energy_conservation}
\end{figure}

To confirm the correctness of the implementation, the energy and gradient were verified against the implementation in PySCF~\cite{sun_recent_2020}. To additionally verify the correctness of the gradient and demonstrate the utility of the software, we perform a 10 ps AIMD simulation at the RI-HF/def2-SVP level on a 
\ce{C54H56} fully hydrogenated diamond nanoparticle in vacuum with 1 fs timesteps. The system was prepared with IQmol and the simulation was started from a non-equilibrium geometry with zero initial velocity such that the total mechanical energy leads to a temperature of approximately 300 K. As shown in Figure~\ref{fig:energy_conservation}, the total energy is conserved demonstrating the correctness of the calculated gradients.

This simulation took 5.65 hours with an average time step latency of 2.03 seconds, thereby enabling a simulation throughput of 42 ps per day. The simulation averaged a performance of 25.2 TFLOP/s which corresponds to 32\% of the theoretical double precision peak on a 4 $\times$ A100 node.

\subsection{Single GPU performance comparison}

\begin{figure}[!h]
    \centering
    \includegraphics[width=1\linewidth]{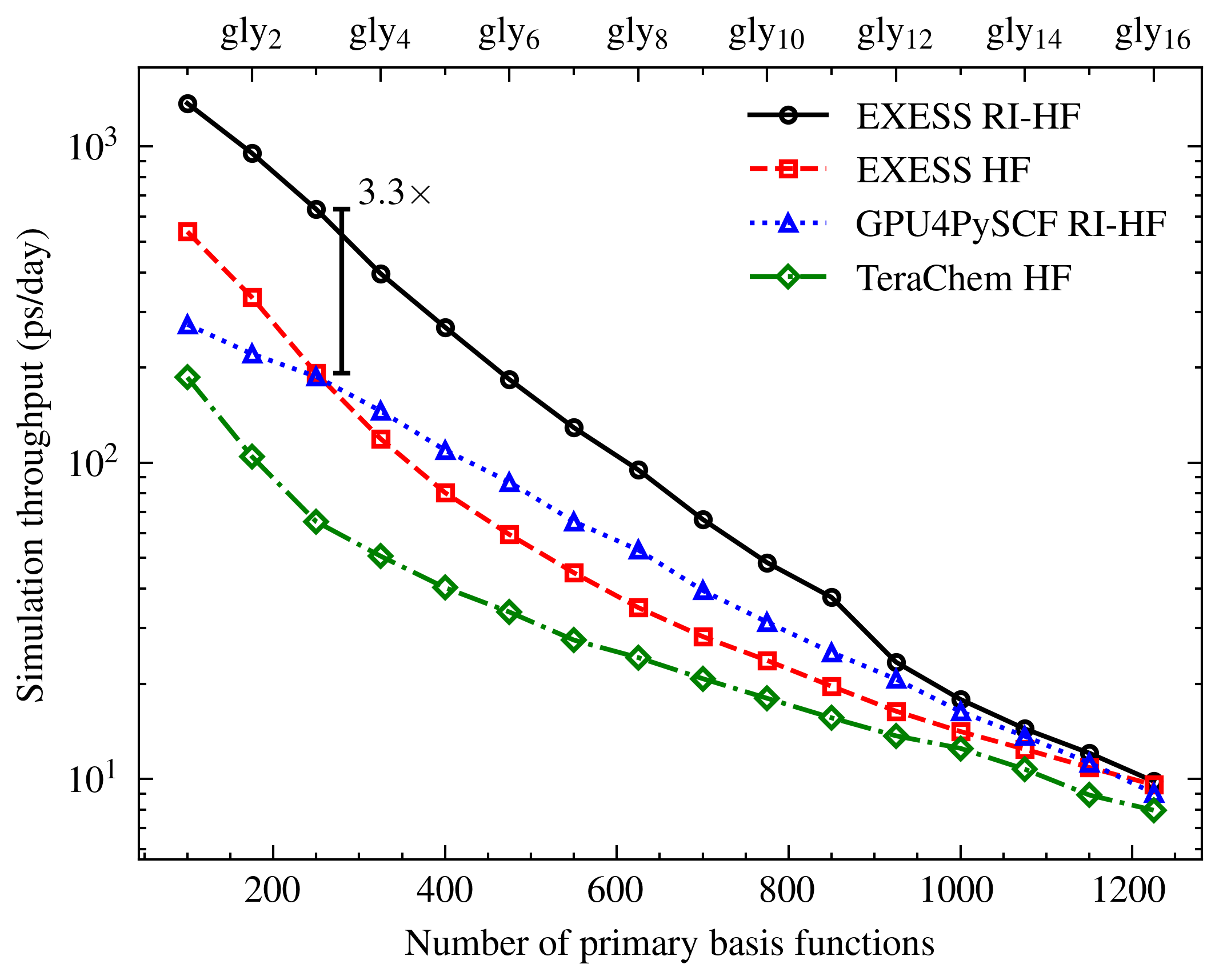}
    \caption{Single GPU HF AIMD throughput for varying length glycine chains with the cc-pVDZ basis set relative to previous RHF implementations in EXESS and TeraChem, and RI-HF implementation in GPU4PySCF. The maximum speedup is labelled. The TeraChem SCF calculation for gly$_15$ failed to converge on time step 20 so only the first 19 time steps are used for the average.}
    \label{fig:perf_compare}
\end{figure}

To analyse the relative performance of the present implementation against existing state-of-the-art software, the single GPU performance was benchmarked against the existing GPU accelerated RI-HF implementation in PySCF as well as the non-RI implementations in EXESS and TeraChem. The results for a range of polyglycine chains are shown in Figure~\ref{fig:perf_compare}. Our implementation demonstrates significant speedup in single-GPU AIMD throughput relative to existing software for systems up to approximately 1000 basis functions, with diminishing returns thereafter.

All software was set up to re-use converged orbitals from the previous time step and convergence thresholds were set such that the total number of SCF iterations were approximately the same between the programs. All programs performed a 100 time step simulation with 1 fs time steps started from a non-equilibrium geometry with zero initial velocity for consistency between the programs.

The drop in throughput from gly$_{11}$ to gly$_{12}$ for the RI-HF implementation is due to the $B_{\mu\nu}^P$ matrix becoming too large to store on the GPU decompressed so the decompression overhead is added to every SCF iteration. This drop occurs for larger systems when utilising multiple GPUs. It would likely be beneficial to allow a slow transition between compressed and decompressed $B$ by storing part of B decompressed if the full matrix does not fit, though this has been left for future implementation.

\subsection{Timing Breakdown}

\begin{figure}
    \centering
    \includegraphics[width=1\linewidth]{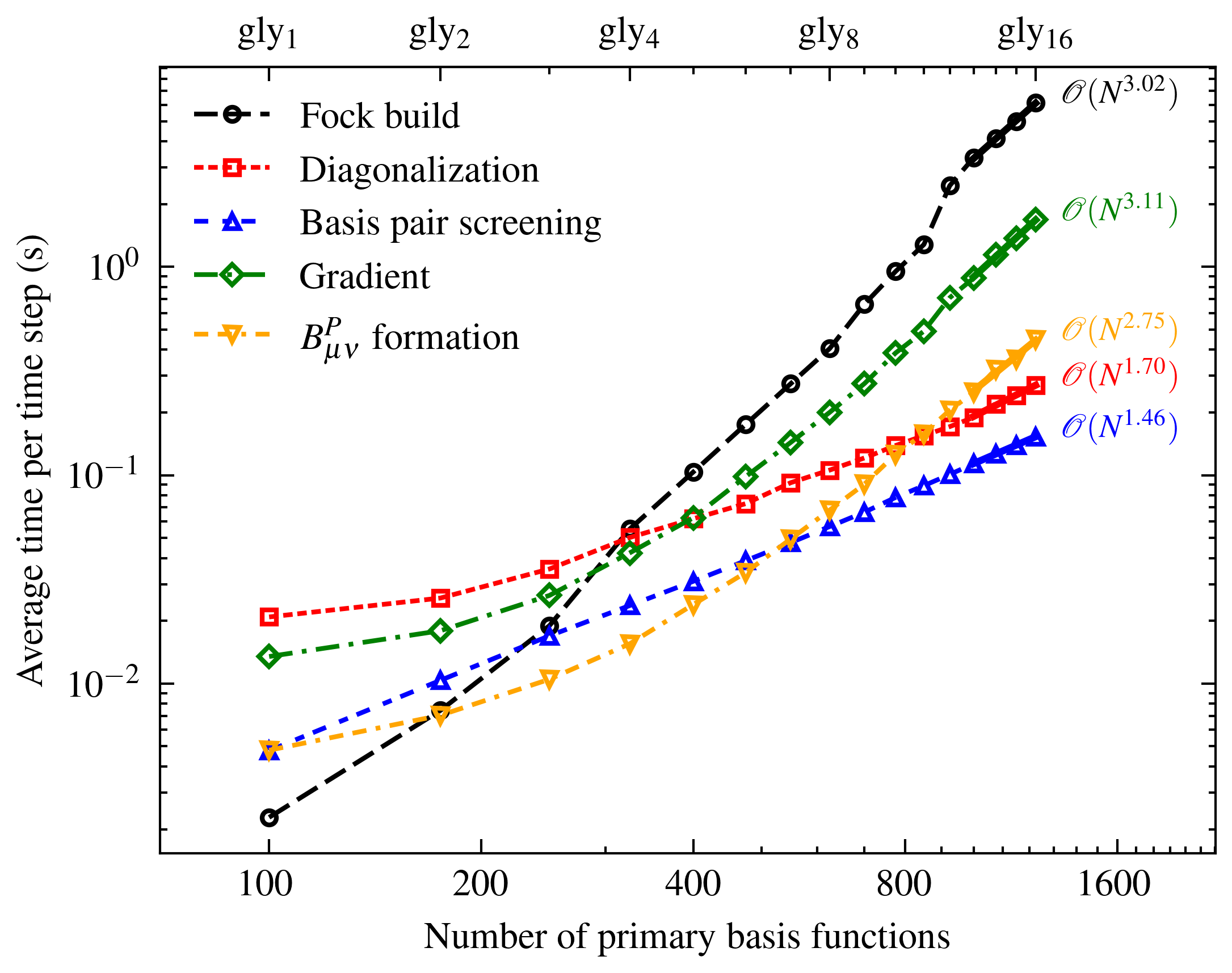}
    \caption{Single GPU timing breakdown of the most significant components of a RI-HF gradient calculation with the cc-pVDZ basis set. The exponent of a polynomial fit to the last four points of each component is labelled. The Fock build component includes decompression time for the systems larger than gly$_{12}$ and the basis pair screening component includes copying the resulting basis function pair data to the GPU.}
    \label{fig:time_breakdown}
\end{figure}

The average time spent in each portion of the RI-HF gradient algorithm per iteration of the AIMD simulation is shown in Figure~\ref{fig:time_breakdown}. Whilst the theoretical scaling is quartic, all components exhibit sub-quartic scaling due to the improved computational efficiency with better screening and higher floating point throughput as the systems get larger.

For small systems (less than 300 basis functions), the diagonalization of the Fock matrix is the dominant component of the execution time. Since  parallelization of the eigenvalue decomposition is extremely inefficient across multiple GPUs, these systems are not expected to experience much speedup from the use of multiple GPUs.

For larger systems, the Fock build becomes the dominant component, taking 70\% of the total execution time for gly$_{16}$. In contrast, just 3\% of the time is spent in the eigenvalue decomposition. 15 to 20\% of the total time is spent in the gradient component regardless of system size.

\subsection{Strong Scaling}

\begin{figure}[!h]
    \centering
    \includegraphics[width=\linewidth]{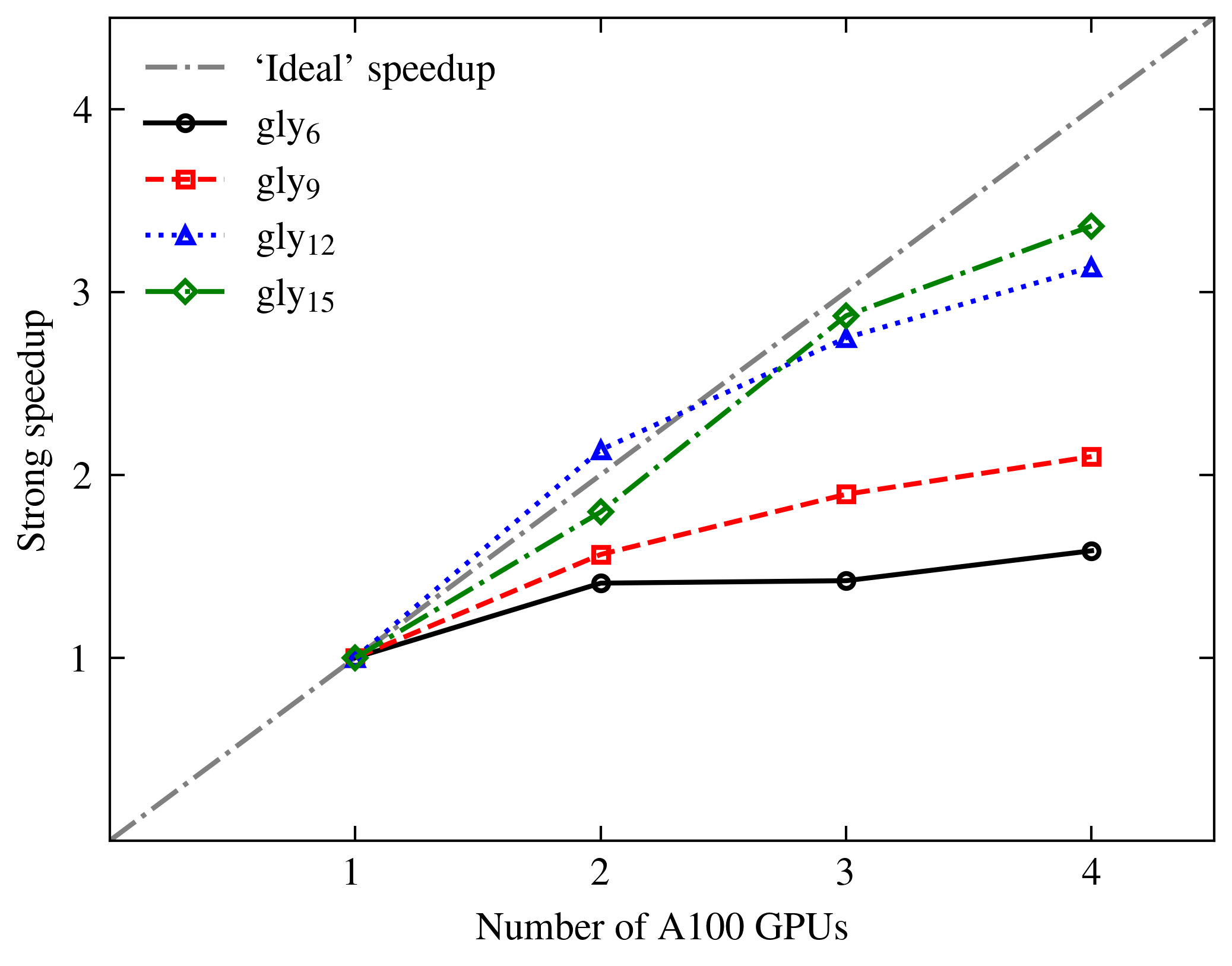}
    \caption{Strong scaling performance of RI-HF AIMD implementation on 1 to 4 A100 GPUs. Tests were conducted on polyglycine chains with the 6-31G**/6-31G**-RIFIT basis sets for 100 1 fs timesteps.}
    \label{fig:strong}
\end{figure}

An important aspect of a multi-GPU implementation is the parallel efficiency relative to using just a single GPU. We conducted a strong scaling analysis for a range of polyglycine chains, as shown in Figure~\ref{fig:strong}.

The strong scaling efficiency is very good for systems where the additional aggregate GPU memory allows the full $B_{\mu\nu}^P$ matrix to be stored decompressed. For the gly$_{12}$ chain this occurs going from 1 to 2 GPUs providing a super linear speedup. For gly$_{15}$ this occurs between 2 and 3 GPUs.

The strong scaling for the smaller gly$_6$ and gly$_9$ systems is significantly lower as the $B_{\mu\nu}^P$ matrix can be stored decompressed on a single GPU and a greater portion of the time is spent in the non-parallelizable overheads for three center integral computation and eigenvalue decomposition of the Fock matrix. Using a full node still provides up to $2\times$ speedup for the gly$_9$ system which could enable longer time scale simulations with lower latency.

\begin{table*}[!t]
\resizebox{\textwidth}{!}{
\begin{tabular}{ccccccccc}
\toprule
\textbf{System} &
  \textbf{Formula} &
  \textbf{Basis} &
  \textbf{\begin{tabular}[c]{@{}c@{}}Basis functions\\ (Primary/Aux)\end{tabular}} &
  \textbf{Compressed} &
  \textbf{\begin{tabular}[c]{@{}c@{}}Time per \\ MD step (s)\end{tabular}} &
  \textbf{\begin{tabular}[c]{@{}c@{}}Average SCF\\ iterations\end{tabular}} &
  \textbf{\begin{tabular}[c]{@{}c@{}}Performance\\ (TFLOP/s)\end{tabular}} &
  \textbf{\begin{tabular}[c]{@{}c@{}}Throughput\\ (ps / day)\end{tabular}} \\ \midrule
\multirow{12}{*}{\textbf{Water}}              & \multirow{3}{*}{$(H_2O)_{10}$}              & def2-SVP  & 250 / 850   & N & 0.12  & 8.97  & 1.32  & 717.2 \\
                                              &                                             & def2-TZVP & 480 / 1270  & N & 0.29  & 9.02  & 3.00  & 298.0 \\
                                              &                                             & def2-QZVP & 1420 / 3330 & N & 2.50  & 10.04 & 9.60  & 34.5  \\ \cline{2-9} 
                                              & \multirow{2}{*}{$(H_2O)_{20}$}              & def2-SVP  & 500 / 1700  & N & 0.32  & 9.01  & 7.44  & 268.6 \\
                                              &                                             & def2-TZVP & 960 / 2540  & N & 0.95  & 9.41  & 12.59 & 90.6  \\ \cline{2-9} 
                                              & \multirow{2}{*}{$(H_2O)_{30}$}              & def2-SVP  & 750 / 2550  & N & 0.66  & 9.02  & 16.41 & 129.8 \\
                                              &                                             & def2-TZVP & 1440 / 3810 & N & 2.50  & 9.95  & 23.41 & 34.6  \\ \cline{2-9} 
                                              & \multirow{2}{*}{$(H_2O)_{40}$}              & def2-SVP  & 1000 / 3400 & N & 1.30  & 9.03  & 25.33 & 66.4  \\
                                              &                                             & def2-TZVP & 1920 / 5080 & N & 5.89  & 9.98  & 29.61 & 14.7  \\ \cline{2-9} 
                                              & \multirow{2}{*}{$(H_2O)_{50}$}              & def2-SVP  & 1250 / 4250 & N & 2.34  & 9.14  & 33.42 & 36.8  \\
                                              &                                             & def2-TZVP & 2400 / 6350 & Y & 12.50 & 9.98  & 33.21 & 6.9   \\ \cline{2-9} 
                                              & $(H_2O)_{60}$                               & def2-SVP  & 1500 / 5100 & Y & 4.77  & 9.5   & 34.13 & 18.1  \\ \hline
\multirow{2}{*}{\textbf{Chlorophyll C2}}      & \multirow{2}{*}{$C_{35}H_{28}MgN_4O_5$}     & def2-SVP  & 819 / 2890  & N & 1.10  & 14.35 & 19.88 & 78.5  \\
                                              &                                             & def2-TZVP & 1787 / 4738 & N & 6.41  & 14.88 & 26.39 & 13.5  \\ \hline
\multirow{2}{*}{\textbf{Cucurbit{[}5{]}uril}} & \multirow{2}{*}{$C_{30}H_{30}N_{20}O_{10}$} & def2-SVP  & 1050 / 3750 & N & 1.71  & 9.16  & 27.45 & 50.6  \\
                                              &                                             & def2-TZVP & 2340 / 6180 & Y & 15.59 & 9.99  & 24.69 & 5.5   \\ \hline
\multirow{2}{*}{\textbf{Bucky ball}}          & \multirow{2}{*}{$C_{60}$}                   & def2-SVP  & 900 / 3300  & N & 1.18  & 8.1   & 21.97 & 73.2  \\
                                              &                                             & def2-TZVP & 2160 / 5700 & Y & 15.72 & 11.3  & 19.90 & 5.5   \\ \hline
\multirow{2}{*}{\textbf{Cholestrol}}          & \multirow{2}{*}{$C_{27}H_{46}O$}            & def2-SVP  & 650 / 2230  & N & 0.50  & 8.9   & 11.09 & 171.8 \\
                                              &                                             & def2-TZVP & 1284 / 3396 & N & 1.96  & 9.58  & 17.51 & 44.0  \\ \hline
\multirow{2}{*}{\textbf{\begin{tabular}[c]{@{}c@{}}Circumcoronene\\ (Graphene)\end{tabular}}} &
  \multirow{2}{*}{$C_{54}H_{18}$} &
  def2-SVP &
  900 / 3240 &
  N &
  1.09 &
  9.78 &
  22.87 &
  78.8 \\
                                              &                                             & def2-TZVP & 2052 / 5418 & Y & 9.98  & 10.79 & 22.72 & 8.7   \\ \bottomrule
\end{tabular}
\caption{Floating-point performance results and timings for a variety of organic chemical systems on 4$\times$A100 GPUs. The compressed column marks whether it was possible to store $B_{\mu\nu}^P$ uncompressed or whether it had to be decompressed each iteration to fit in GPU memory. }\label{tab:timings} 
}\end{table*}

\subsection{Efficiency}

\begin{figure}
    \centering
    \includegraphics[width=1\linewidth]{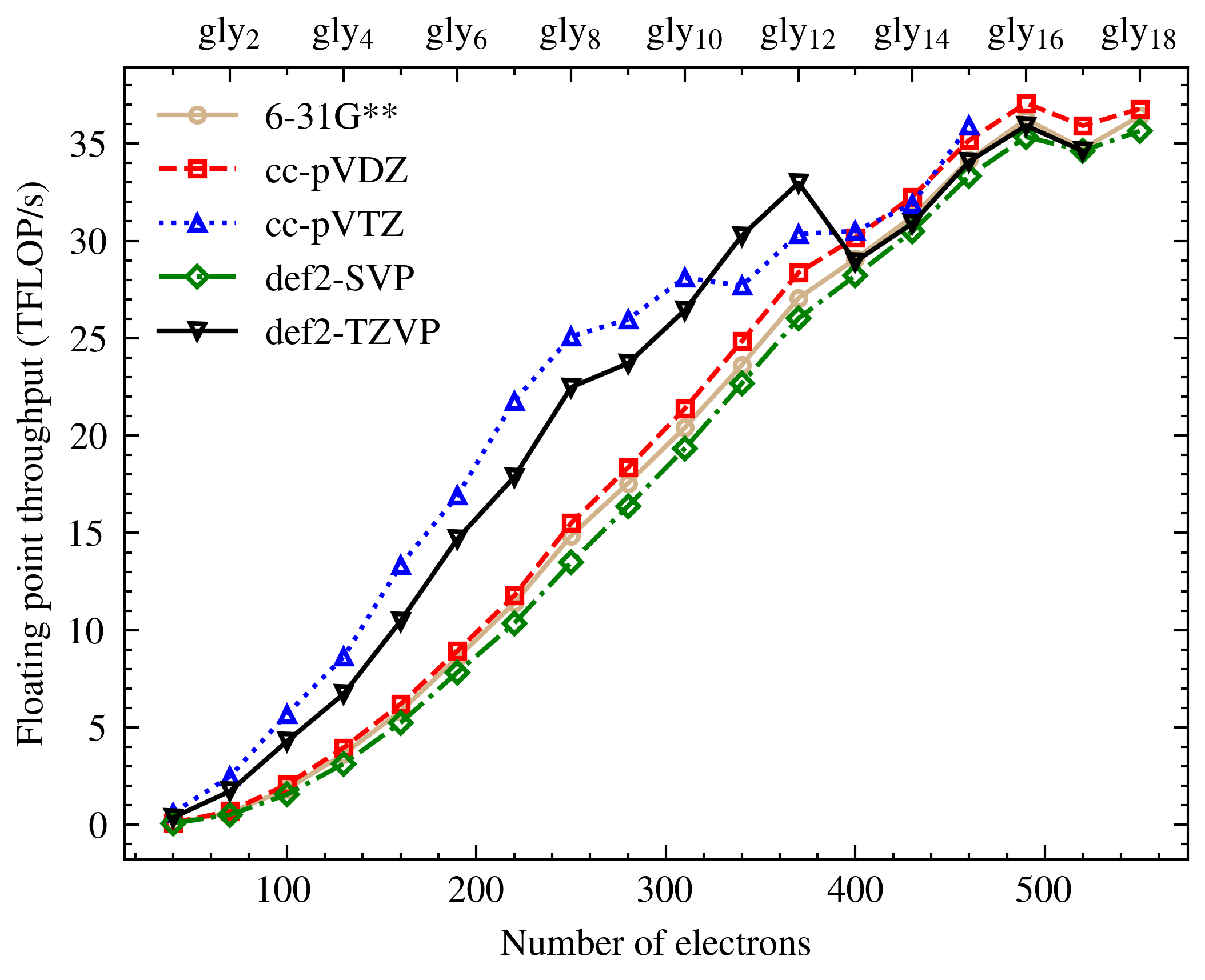}
    \caption{Average floating point rate for various polyglycine chains and basis sets on a 4 $\times$ A100 node.}
    \label{fig:efficiency}
\end{figure}

To investigate the hardware utilization efficiency of the RI-HF implementation we measured the average floating point throughput for a variety of system sizes and basis sets as shown in Figure~\ref{fig:efficiency}.

The triple zeta basis sets provide greater floating point throughput for smaller systems due to the larger number of basis functions producing larger matrices. The drop in performance from gly$_{12}$ to gly$_{13}$ for def2-TZVP is due to running out of GPU memory on the 4 $\times$ 80 GB GPUs and having to store $B_{\mu\nu}^P$ compressed. 

The maximum achieved full calculation performance on 4 A100 GPUs was 37 TFLOP/s for the gly$_{16}$ chain with cc-pVDZ which corresponds to 47\% of the theoretical peak performance. However for these large systems, the benefit of the RI approximation is significantly diminished due to the higher formal scaling as shown in Figure~\ref{fig:perf_compare}. The primary limitations for the smaller systems are the matrix dimensions producing inefficient GEMM calls and time spent in the cubic scaling integral routines and eigenvalue decomposition.

An array of additional performance timings comparing double zeta and triple zeta basis sets is provided in Table~\ref{tab:timings}. These timings were all obtained using all four A100 GPUs on a Perlmutter node using a 100 $\times$ 1 fs time step simulation to amortize startup effects. Whilst 1 fs is unnecessarily small for the \ce{C60} system without any hydrogens, 1 fs is used for consistency and performance comparison (a higher ps / day throughput could be obtained by increasing the time step without introducing significant error as the higher mass atoms move much slower than the hydrogens). A conservative convergence threshold of $10^{-10}$ Hartree for the sum of the absolute values of the DIIS error vector is used to ensure accurate gradients in all tests.

\subsection{Fragmented AIMD: A Proof of Concept}

In this subsection, we demonstrate a proof-of-concept AIMD simulation as a potential application of the improved performance for small and medium size molecules provided by the present RI-HF implementation.

We performed a 1 ps simulation with 1 fs time steps on a segment of the 6PQ5 prion-protein with 1,380 electrons across 360 atoms. The system was split into 36 small fragments with 7 to 14 atoms each. An MBE-3 expansion and the cc-pVDZ/cc-pVDZ-RIFIT basis sets were utilized. The dimer and trimer cutoffs for maximum distance between constituent monomer centroids were $22$ {\AA} and $9$ {\AA} respectively.

The simulation was conducted on 128 nodes of the Perlmutter supercomputer, with two worker teams per node such that each fragment is assigned to 2 $\times$ A100 GPUs. The calculation averaged 0.76 seconds per time step which corresponds to a simulation throughput of 113 ps/day, potentially enabling full quantum nanosecond simulations of a protein-scale system.

This simulation is an indicative proof-of-concept and full accuracy and maximal performance benchmarks have not yet been performed. It is not expected to conserve energy at this stage due to fragments moving in and out of the cutoff radius during the simulation.

Orbitals were not re-used as an initial guess for each time-step due to the dynamic distribution of fragments to nodes so significant time is spent in the first SCF iteration performing the RI Fock build from a density guess. Future work is planned to perform an orbital initial guess to eliminate this overhead. We also intend to improve the accuracy further using Hybrid-DFT techniques and correlation from post-HF corrections such as MP2.

\section{Conclusions}

In this Article, we presented an optimized multi-GPU algorithm for the calculation of RI-HF energies and analytic gradients, specifically tailored for high throughput AIMD simulations. Our approach leverages several key innovations, including multi-GPU parallelism, efficient workload balancing, and utilization of symmetry and sparsity to enhance computational efficiency and memory usage.

Our results demonstrate significant improvements in performance compared to previous GPU-accelerated RI-HF and traditional HF methods, with single GPU speedups exceeding 3 $\times$. Additionally, the use of multiple GPUs can achieve super-linear speedups by taking advantage of the increased aggregate GPU memory, which allows for the storage of decompressed three-center integrals. This capability is particularly beneficial for larger molecular systems, as it reduces the need for repeated decompression and thus lowers computational overheads.

Practical applications of our algorithm were demonstrated through extensive performance benchmarks using up to quadruple-$\zeta$ primary basis sets. Notably, the implementation achieved up to 47\% of the theoretical peak floating-point performance on a 4$\times$A100 GPU node, showcasing its effectiveness in utilizing modern parallel computing architectures.

We validated the correctness of our energy and gradient calculations against existing implementations and demonstrated the practical utility of our approach through AIMD simulations. For instance, in a 10 ps simulation of a fully hydrogenated diamond nanoparticle, our method ensured energy conservation and achieved a simulation throughput of 42 ps per day, highlighting its suitability for molecular dynamics studies.

Our work also includes a proof-of-concept demonstration of a fragmented AIMD simulation on a prion-protein segment, which suggests the potential for full quantum nanosecond simulations of protein-scale systems. This capability, combined with the efficient handling of smaller molecular fragments, opens up new avenues for high-accuracy, larger time-scale molecular simulations.

\section{Acknowledgements}

The authors thank the National Energy Research Scientific Computing Center (NERSC), a Department of Energy Office of Science User Facility using award ERCAP0026496 for resource allocation on the Perlmutter supercomputer.
RS and EP thank QDX technologies for providing additional funding to support the development of the software.

\bibliography{used_refs}

\end{document}